\documentclass{article}

\usepackage{arxiv}

\usepackage[utf8]{inputenc} 
\usepackage[T1]{fontenc}    
\usepackage{hyperref}       
\usepackage{url}            
\usepackage{booktabs}       
\usepackage{amsfonts}       
\usepackage{nicefrac}       
\usepackage{microtype}      
\usepackage{lipsum}
\usepackage{graphicx}
\graphicspath{ {./images/} }

\title{Boosted decision trees approach to neck alpha events discrimination in DEAP-3600 experiment}

\author{
 Ilyasov Aidar \\
  National Research Center ``Kurchatov Institute''\\
  Kurchatov sq. 1, Moscow, Russia\\
  National Research Nuclear University MEPhI \\
  Kashirskoe sh. 31, Moscow, Russia \\
  \texttt{ilyasovaid@yandex.ru} \\
   \And
 Grobov Alexey \\
  National Research Center ``Kurchatov Institute''\\
  Kurchatov sq. 1, Moscow, Russia\\
  National Research Nuclear University MEPhI \\
  Kashirskoe sh. 31, Moscow, Russia \\
  \texttt{alexey.grobov@gmail.com} \\
}

\begin{document}
\maketitle
\begin{abstract}
Machine learning (ML) has been widely applied in high energy physics to help the physical community in particle classification and data analysis. Here we describe the application of machine learning to solve the problem of classifying background and signal events for the DEAP-3600 dark matter search experiment (SNOLAB, Canada). We apply Boosted Decision Trees (BDT) algorithm of ML with improvements from Extra Trees and eXtra Gradient Boosting (XGBoost) methods \cite{chen2016xgboost} \cite{amaudruz2019design} .\\
\end{abstract}


\section{Introduction}
A general problem in experimental high energy physics is the large amount of data from detectors and, in accordance with this, labor of time and human resources for data processing. Some of these problems can be solved with machine learning (ML). The advantages of ML are high performance and ability to find hidden patterns in data.

DEAP-3600 (Dark matter Experiment using Argon Pulse-shape discrimination) is a single phase liquid argon dark matter experiment, located 2 km underground in Canada. The design cross section sensitivity for DEAP-3600 to the spin-independent scattering of Weakly Interacting Massive Particles (WIMPs) on nucleons is $10^{-46}$ $cm^2$ for a 100 GeV/$c^2$ WIMP mass with a fiducial exposure of 3 tonne-years \cite{amaudruz2019design}.

Given the sensitivity of the detector, it is natural to have a very thorough event selection process. Cut-based analysis proved to be very effective in most cases, but insensitive to non-linear dependencies in data. This is where machine learning goes on stage. Here we describe training and testing process of a boosted decision tree (BDT) classifier, techniques which were used to improve the accuracy of the algorithm and results of the approach to the problem of neck alpha background mitigation.

The use of cut-based analysis leaves us with 20\% of the signal acceptance, with a background level less than 0.5 events. Here we report increase of signal acceptance up to 48\% of signal acceptance at the background rejection rate of 99.9\%.

\section{DEAP-3600 experiment}
The DEAP-3600 experiment has the following characteristic features\cite{collaboration2019search}:

\begin{figure}[h]
    \centering
    \includegraphics[scale=0.3]{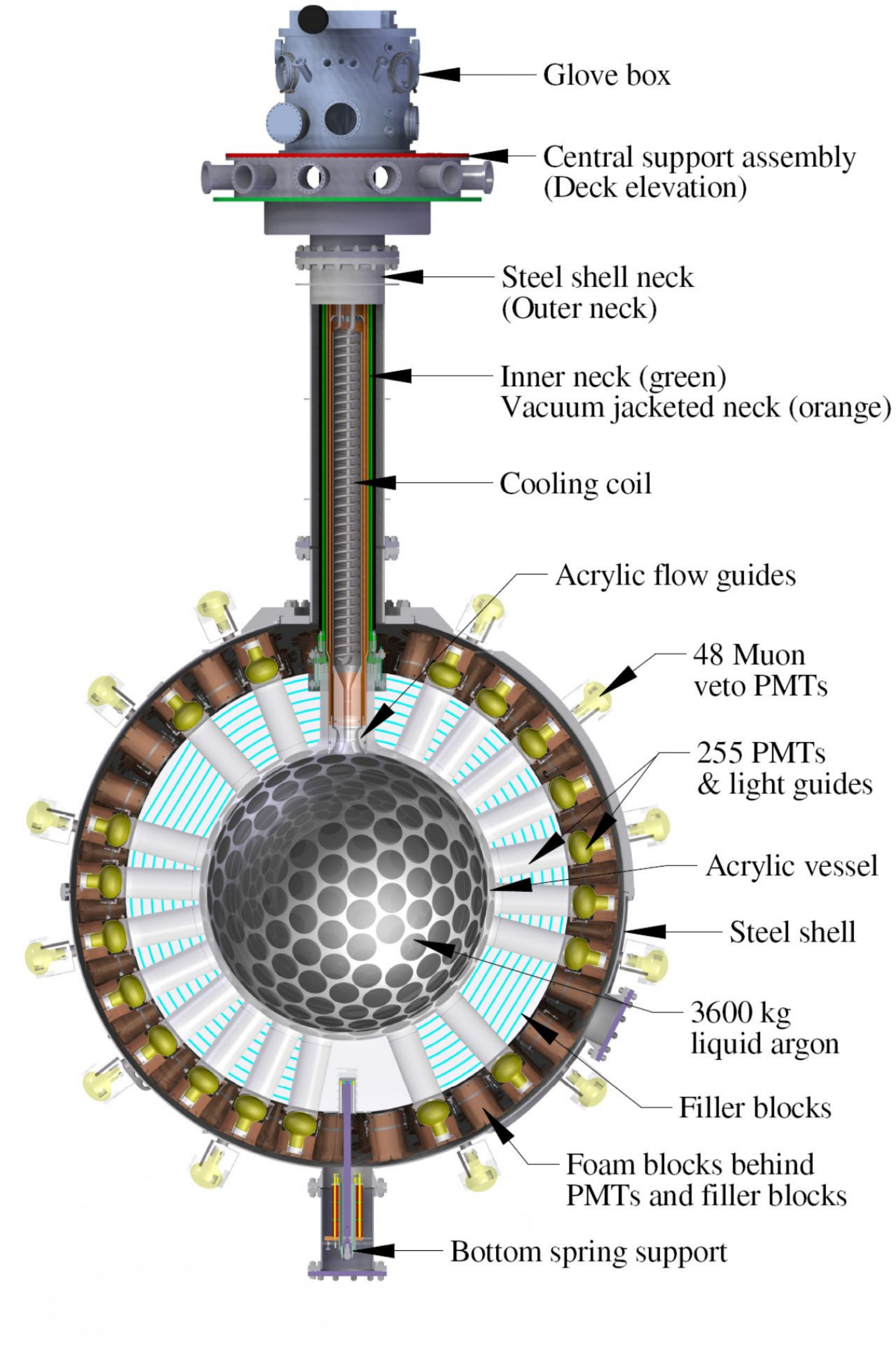}
    \caption{DEAP-3600 detector scheme \cite{collaboration2019search}.}
    \label{ris:image}
\end{figure}

\begin{enumerate}
    \item Located approximately 2km underground;
    \item Limit on the WIMP-nucleon spin-independent cross section on a liquid argon (LAr) target of 3.9 $\times$ 10$^{-45}$ cm$^2$  (1.5 $\times$ 10$^{-44}$ cm$^2$) for a 100 GeV/c$^2$ (1 TeV/c$^2$ ) WIMP mass at 90\% confidence level (C.L.) \cite{collaboration2019search};
    \item It consists (3279 $\pm$ 96) kg of LAr with 30 cm of gaseous argon (GAr) at the top;
    \item The argon is contained in a 5 cm thick ultraviolet absorbing (UVA) acrylic vessel (AV) with an inner diameter of 1.7 m;
    \item The GAr and LAr regions are viewed by an array of 255 inward-facing 8" diameter Hamamatsu R5912 HQE low radioactivity photomultiplier tubes (PMTs);
    \item The detector has neutron shielding:
    \begin{enumerate}
        \item 45 cm light guides (LGs) between PMTs and AV;
        \item Space between LGs has filler blocks (FBs).
    \end{enumerate}
    \item The top opens to the neck with liquid $N_{2}$ -filled cooling coil, used to fill the detector;
    \item Optical fibers wrap around base of neck, coupled to 4 PMTs -- neck veto (NV);
    \item All enclosed in stainless steel shell (SSS);
    \item The detector has a Cherenkov muon veto:
    \begin{enumerate}
        \item Submerged in 300 tonnes $H_{2}O$;
        \item 48 outward-facing PMTs on SSS.
    \end{enumerate}
\end{enumerate}
One of the biggest contribution to the background rate for the WIMP search is $\alpha$ particles. Signals from $\alpha$-decays from short- and long-lived
$^{222}Rn$ progeny as well as short-lived $^{220}Rn$ progeny are observed at several locations inside the detector.
\begin{figure}[h]
\includegraphics[width=1\textwidth]{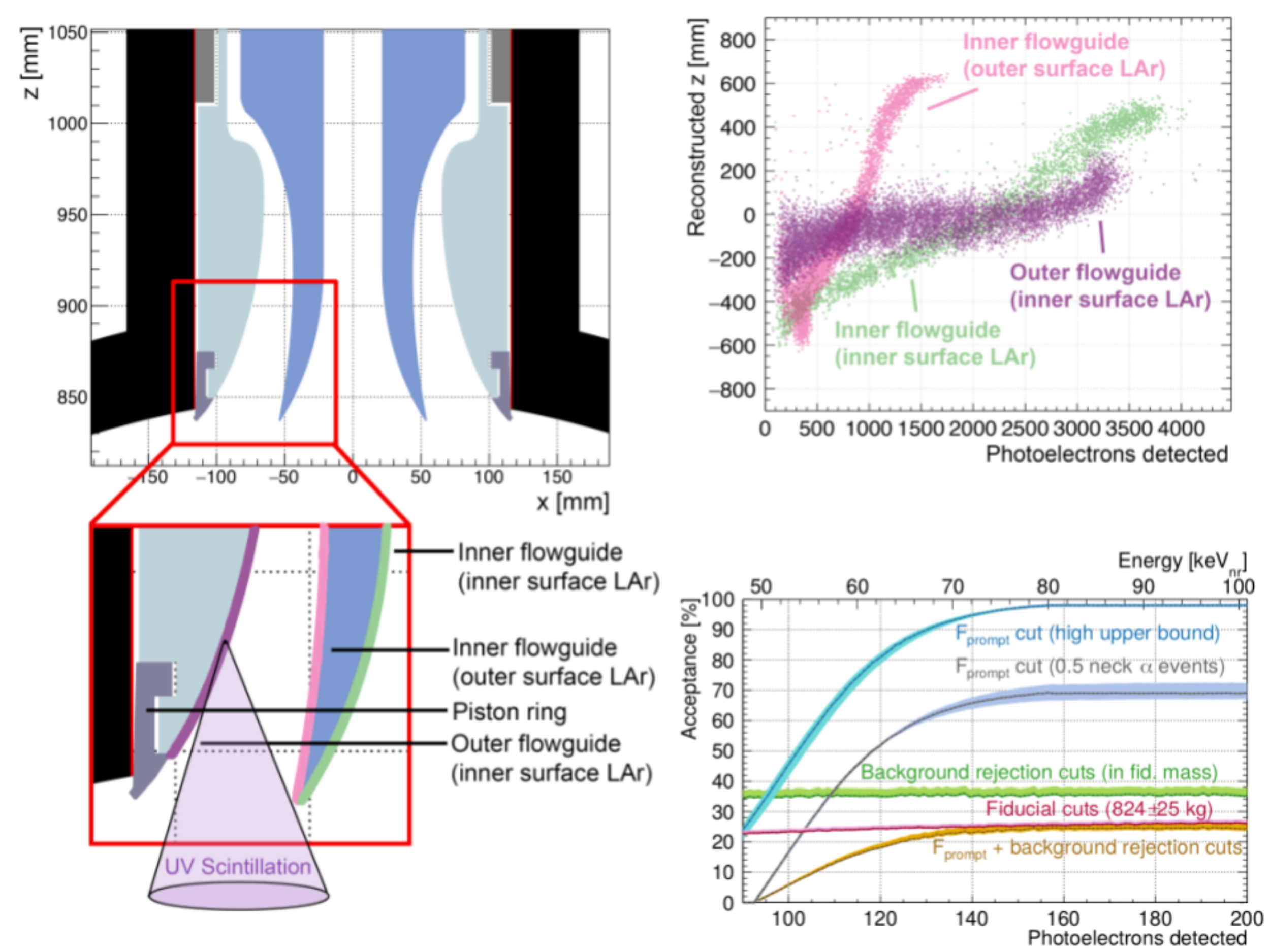}
\caption{Left: Cross-sectional illustration of the FG components in the AV neck. Shown are the three FG surfaces and the piston ring (not coated in LAr for purposes of illustration). Top right: Simulated relationship in reconstructed z vs. number of photoelectrons (PE) for $\alpha$-decays on the IFG-IS (green), IFG-OS (pink), and OFG-IS (purple). Bottom right: WIMP acceptance as a function of PE, broken down by cut type\cite{collaboration2019search}.}
\label{alphas}
\end{figure}
These include the LAr target, the LAr/TPB and TPB/AV (here TPB - 3 $\mu$m layer of wavelength shifter, 1,1,4,4-tetraphenyl-1,3-butadiene) surfaces, and the surfaces of the acrylic flowguides (FGs) in the AV neck. After applying fiducial cuts, the largest contribution to the background rate is from $^{210}Po$ $\alpha$-decays on the surfaces of the acrylic  FGs in the AV neck \cite{collaboration2019search}. There are three distinct surfaces on these components: the inner flowguides inner and outer surfaces (IFG-IS and IFG-OS) and the outer flowguides inner surface (OFG-IS, Figure \ref{alphas}). The $\alpha$-decays on the piston ring have a negligible contribution to the background rate in the WIMP region-of-interest (ROI) \cite{collaboration2019search}. 

Two main cuts have been developed such that the WIMP acceptance is maximized while maintaining a background expectation below the benchmark < 0.5 events from all flowguide components:
\begin{enumerate}
    \vspace{-0.2cm}
    \item Upper $F_{prompt}$ cut (Figure \ref{alphas}). $F_{prompt}$ is a pulseshape parameter and defined as follows \cite{collaboration2019search}:
    \begin{equation}
        F_{prompt} \equiv  \frac{\sum_{{i|t_{i} \in (-28 ns, 60 ns)}}Q_{i}}{\sum_{{i|t_{i} \in (-28 ns, 10 \mu s)}}Q_{i}}
        \label{fprompt}
    \end{equation}
    This cut removes a significant fraction of $\alpha$-decays in the neck, at the cost of 30\% acceptance loss to signal events \cite{collaboration2019search}.
    \vspace{-0.2cm}
    \item Early pulses in GAr PMTs. This cut removes neck $\alpha$-decays where the scintillation light reflects off the GAr/LAr interface and hits a PMT in the gaseous volume very early. Event are rejected if any of the first 3 pulses observed in the PE integration window are registered in PMTs that subtend the GAr region. In this case we reach the neck $\alpha$-decay background target of < 0.5 events when combined with all other cuts. With a 49\% signal acceptance, this cut is the largest source of loss in signal acceptance for events passing low-level and veto cuts within the fiducial volume \cite{collaboration2019search}.
\end{enumerate}

\begin{figure}[h]
    \centering
    \includegraphics[width=0.8\linewidth]{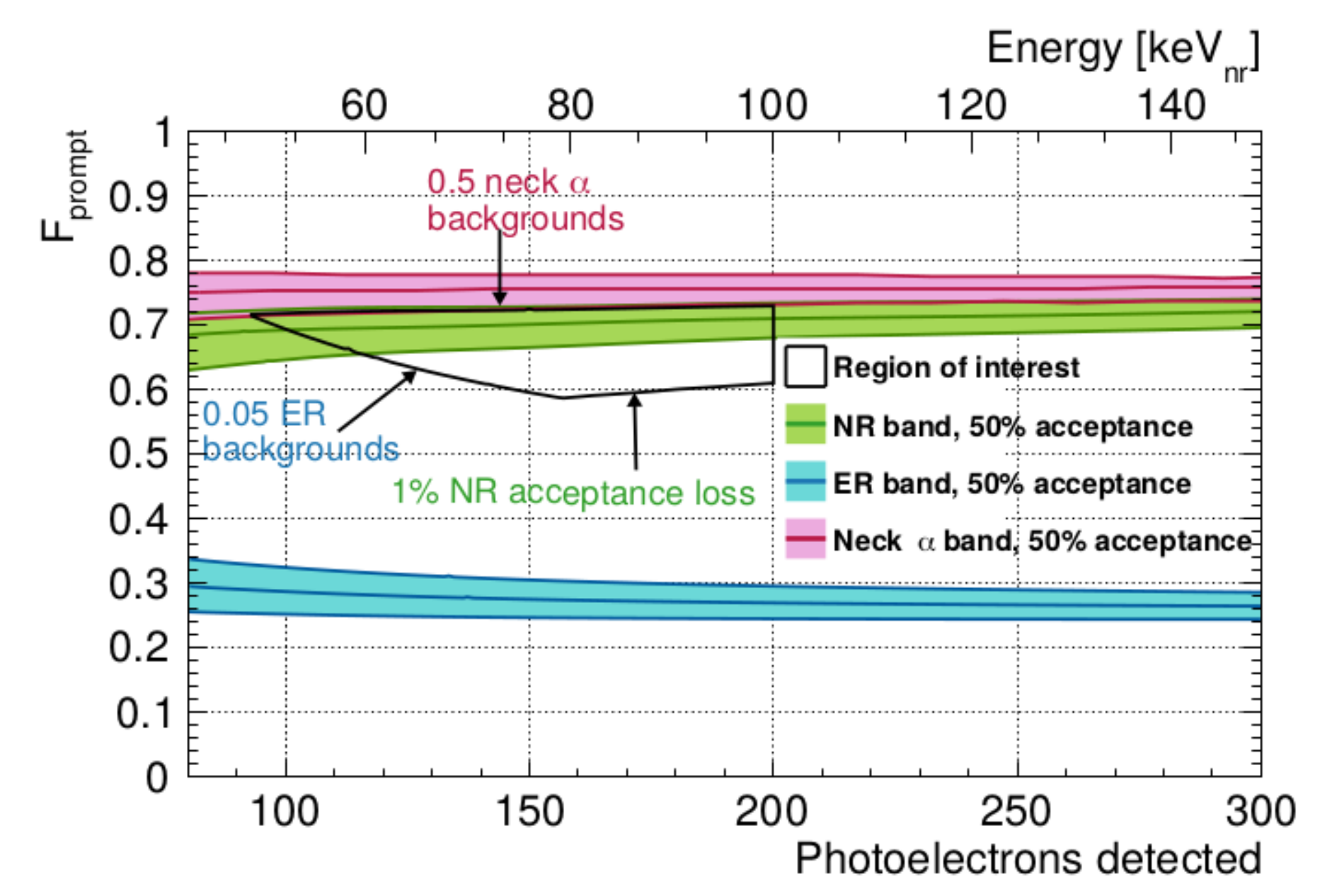}
    \caption{Illustration of the WIMP ROI (black) along with the electronic recoils (ER, blue), nuclear recoils (NR, green) and neck $\alpha$-decay (pink) bands that define the boundaries \cite{collaboration2019search}.}
    \label{ROI}
\end{figure}
After applying all fiducial and background rejection cuts $0.07_{-0.07}^{+0.13}$, $0.17_{-0.14}^{+0.12}$ and $0.25_{-0.20}^{+0.21}$ events
from the IFG-IS, IFG-OS and OFG-IS components are expected in the WIMP ROI, respectively. This combines to an overall expectation of $N_{\alpha, neck}^{ROI} = 0.49_{-0.26}^{+0.27}$ events in the dataset. Figure \ref{ROI} illustrated the WIMP ROI with ER, NR and neck $\alpha$-decay bands. Each band is drawn about the median of each class of event, with 25\% of such events above and 25\% below the shaded regions. So the cut analysis efficiently removes neck $\alpha$ events, but it gives a severe reduction in WIMP acceptance. To improve this we use machine learning.

\section{Boosted Decision Tree method}
\begin{figure}[h]
    \centering
    \includegraphics[width=0.5\linewidth]{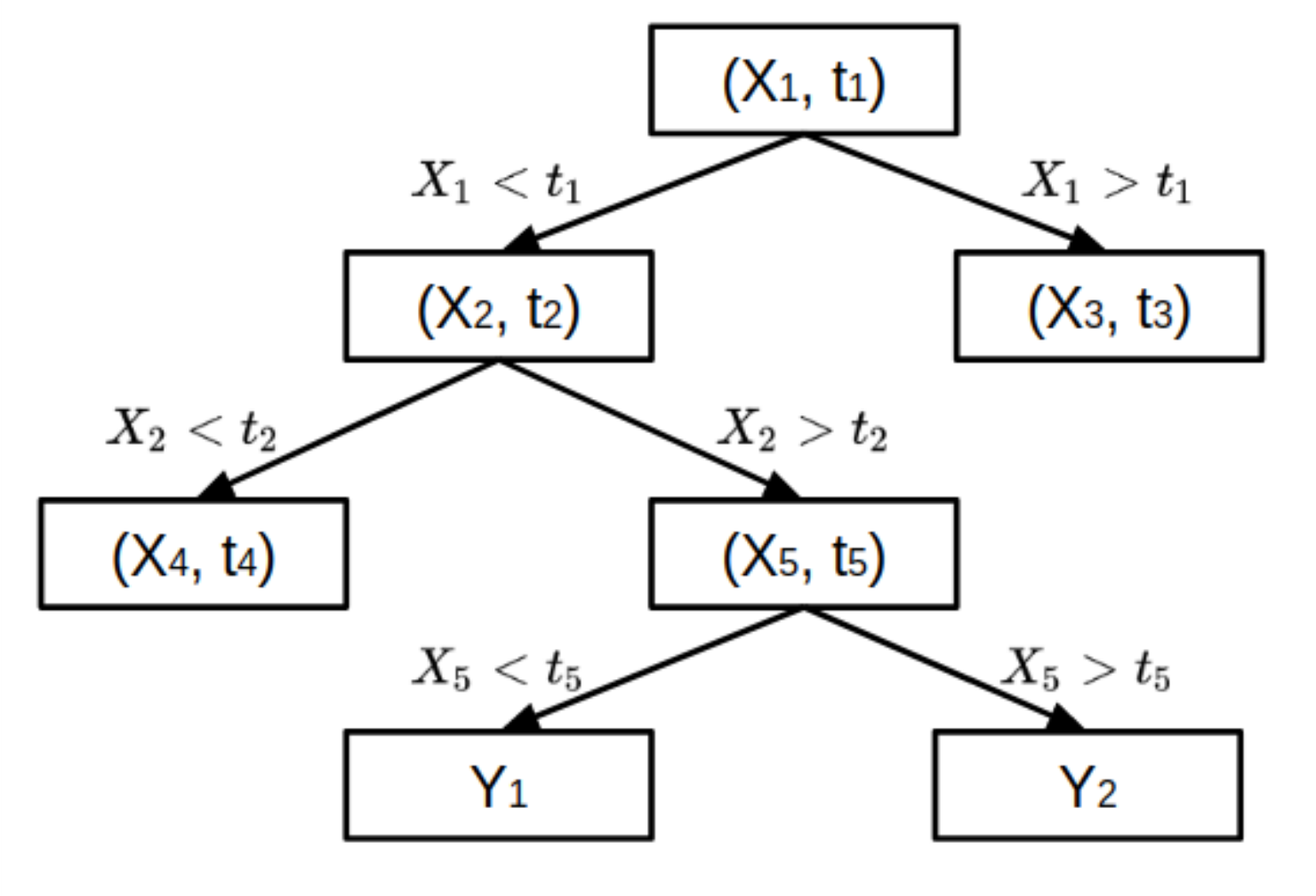}
    \caption{The structure of decision tree: $x_{1-5}$ - features of events, $t_{1-5}$ - threshold values, $Y$ - output labels}
    \label{tree_structure}
\end{figure}

Decision trees have been widely applied in high energy physics with powerful results \cite{albertsson2018machine}.  It is an algorithm that represents a set of rules by which decisions are made. Graphically it looks like a tree structure (Figure \ref{tree_structure}). The main advantage of this model is the simplicity of understanding its calculations (``white box''), but it works better with a gradient boosting method, since boosted trees not inclined to overtrain as much as usual trees. In general we have number of instances $X = ( x_i )$ with actual class labels $Y= ( y_i )$, $y_i = 0$(background) or $1$(signal). This is usually called training sample. Each instance in the training sample has number $N$ of features $x_i (x_{i1}, x_{i2},...,x_{ij})_{j=N}$. Based on these features we try to make a prediction $\bar{y}_i = \sum_j w_j x_{ij}$ for instance $x_i$ to belong to class 0 or 1, where $w_j$ is weight we assign to features of instances. The more the weights the more important the feature. We can compare output with actual label $y_i$, and tune our predictive model. For that we use objective function:
\begin{equation}
    Obj = L(y,\bar{y}) + R(w),
\end{equation}
where $L(y,\bar{y})$ is loss function and $R(w)$ is a regularization function to prevent overfitting on training sample. Decision trees learn by comparing the value of the feature with a threshold value. The first node asks if $x_{11}$ is less than some threshold $t_{1}$. If yes, we then ask if $x_{12}$ is less than some other threshold $t_{2}$. If not, we then ask if $x_{15}$ more than some third threshold $t_{5}$. And depends on results of that questions we get class label \cite{murphy2012machine}. 

Improvement for standard decision trees comes from bagging algorithm: procedure to aggregate the outputs of several trees for voting. This technique is exploited by Random Forest models with minor changes, such as random selected feature subset. The final algorithm is complicated but robust and has a good predictive power. Another enhancement comes from boosting technique: single trees are sequentially built and performance estimation of following trees depends on previous. This helps to assign higher weights to trees with higher performance. Gradient boosting is an advanced technique that exploits gradient descent algorithm to minimize the errors of trees and find a way to quickly converge on the output. 

XGBoost is a model, that combines number of trees into a single effective model in an iterative fashion, effectively using bagging and gradient boosting. It starts from the constant algorithm and adds new ones (trees) while minimizing the objective function. To do that model utilize gradient descent. XGBoost grows trees greedy: introducing new leaf it compares gain and complexity of this step. After tree is grown it is added to the model.

XGBoost enables tree-pruning, thus enhancing productivity and preventing model from overfitting: pruning removes all leafs with negative gain from the model. 

In our case we used XGBoost classifier with logistic regression loss function for binary classification and L1 regularization. 

To find the best classifier parameters, we applied an grid search technique. To use this one must select classifier parameters and the values in which one is interested. This way one defines a hyperparameter grid. This technique will check all possible combinations of parameters on train sample and will display the best. It is very useful, but it requires a lot of execution time to check a large number of input variables.

We adopted following parameters (the bold text marks the parameters that were later used in the analysis):
\begin{enumerate}
    \item \textit{max depth} = (2, 4, 6, 8, \textbf{10}, 12, 14) -- maximum depth of the tree;
    \item \textit{min child weight} = (1, 2, 3, 4, \textbf{5}, 6) -- minimum sum of weights of all observations necessary to create a child object;
    \item \textit{subsample} = (0.5, 0.6, \textbf{0.7}, 0.8, 0.9, 1) -- the proportion of training sample objects used at each iteration (The number of features to consider when looking for the best split);
    \item \textit{colsample by tree} = (0.5, 0.6, 0.7, 0.8, 0.9, \textbf{1}) -- proportion of variables used at each iteration (The minimum number of samples required to split an internal node);
    \item \textit{gamma} = (\textbf{0}, 0.1, 0.2, 0.3, 0.4, 0.5) -- minimum decrease in the value of the loss function.
\end{enumerate}

Note that checking all the parameters at once takes a very large amount of time. However knowing how parameters depend on each other and adjusting dependent parameters separately one can significantly speed up the process. Thus, \textit{max depth} and \textit{min child weight} parameters are configured first. These hyperparameters are recommended to be set together, the higher \textit{max depth} the more complex the model (likelihood of retraining increased as well). At the same time \textit{min child weight} acts as a regularization parameter. Stability of model performance strongly depends on \textit{colsample by tree} and \textit{subsample} parameters.

To quantify how well we identify signal and background events, we define correctly identified signals as ``true positives'' (TPs) and correctly defined backgrounds as ``true negatives'' (TNs). We also define type I error as ``false positives'' (FPs) and type II errors as ``false negatives'' (FNs). The following metrics were used to evaluate the final classifier:\\
\\
    $Accuracy = \frac{TP+TN}{TP+TN+FP+FN}$,
    $Precision = \frac{TP}{TP+FP}$,
    $Recall = \frac{TP}{TP+FN}$,
    $F_{1} - score = 2*\frac{Precision*Recall}{Precision+Recall}$\\
    \\
The classifier was also evaluated using receiver operating characteristics curves (ROC-curves). The ROC-curve is a graph showing the relationship between signal acceptance SA (which is equal to ``recall'') and background rejection rate BRR:\\
\\
$SA = \frac{TP}{TP+FN},\quad BRR = \frac{TN}{FP+TN}$\\
\\
Its quantitative characteristic is the area under the curve (AUC), which equals to ``1'' if our algorithm has an error-free classification. 

\section{Analysis process}

Here, we apply the BDT model to distinguish $^{40}Ar$ from $\alpha$-decay events. 

Feature selection was performed for analysis:
\begin{enumerate}
    \item We deliberately excluded energy estimators from the analysis, since we do not want model to pick up this pattern;
    \item We excluded features defining pulse-shape (equation \ref{fprompt}), $^{40}Ar$ and neck-alpha events share this signature. From the other hand we want to be able to validate our performance on $^{39}Ar$, which has different $F_{prompt}$ parameter;
    \item We included position parameters (reconstructed X,Y,Z coordinates, drift time e.t.c.): given the geometry of the detector it is more likely to neck alpha event to reconstruct near the bottom of the acrylic vessel;
    \item We also included charge distribution patterns and number of hits in different PMTs. We are aware of some correlation between those and position features;
    \item Since background process takes place in the neck we introduce special feature: PMT, which sees the first light in the gaseous phase. 
\end{enumerate}

We used Monte Carlo simulated data samples for both types of events and deliberately avoided variables that contain pulse shape or energy information since we don't want the model to be trained on differences in energy spectra. The course of our work was as follows:

\begin{enumerate}
    \item Shuffle the datasets to avoid overtraining and divide them in a ratio of 70 to 30 into training and testing samples;
    \item Apply machine learning algorithm;
    \item Check the efficiency of each input variable and use feature selection to select the set of parameters that gives the greatest contribution to the result;
    \item Parameters of the model were determined with an exhaustive search procedure;
    \item Check results using quality metrics and ROC-curve.
\end{enumerate}
We tested several algorithms, like gaussian naive bayes, linear classifier (with 
stochastic gradient descent training), decision trees from \textit{python scikit-learn} package and boosted decision trees from \textit{XGBoost} package. \\

\begin{tabular}{ | p{100pt} | r | c | c | c |}
\hline
Algorithm & Accuracy & Precision & Recall & $F_{1}$-score  \\ \hline
BDT & 0.93 & 0.94 & 0.95 & 0.94 \\ \hline
Naive Bayes & 0.86 & 0.91 & 0.84 & 0.87 \\ \hline
Decision tree & 0.88 & 0.90 & 0.90 & 0.90 \\ \hline
Linear classifer & 0.87 & 0.85 & 0.95 & 0.89 \\ 
\hline
\end{tabular}
\\

Best result is achieved with boosted decision trees of \textit{XGBoost}, which proves to be robust and efficient. We choose it to be our benchmark algorithm.

\begin{figure}[h]
    \centering
    \includegraphics[width=0.8\linewidth]{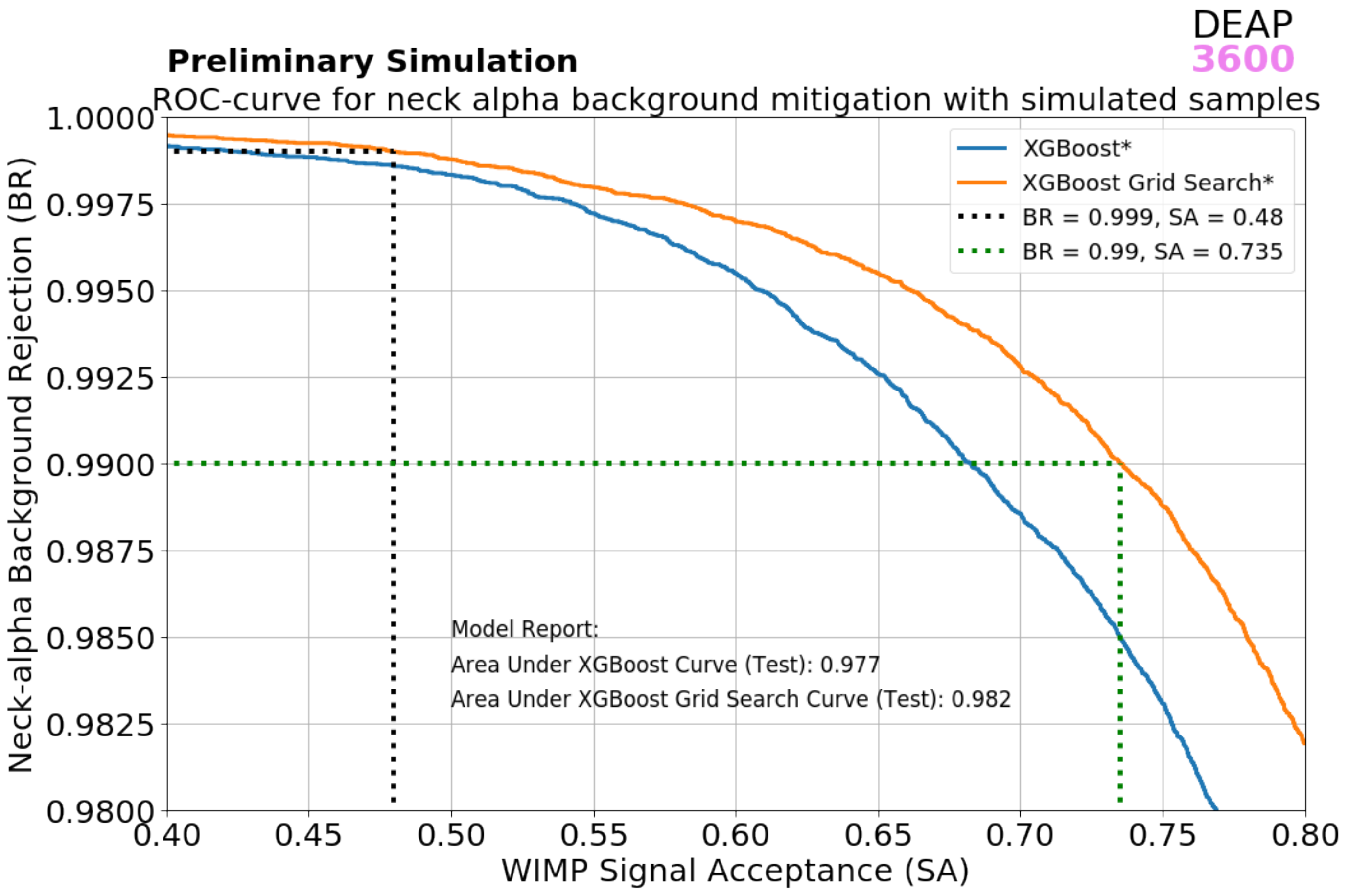}
    \caption{Result of classifier application: relationship between WIMP Signal Acceptance and Neck-alpha Background Rejection with simulated samples. Blue - checking the operation of the classifier without any optimization, orange - the result of the classifier after optimizing the parameters using the grid search technique}
    \label{ROC_AUC}
\end{figure}

Figure \ref{ROC_AUC} shows the main result of this work: the ROC-curve and the AUC-score of the BDT classifier, with and without optimizing parameters using an exhaustive search technique. This graph shows what fraction of simulated WIMP particles after a cut based on the classifier value has been applied, removing some specified fraction of $\alpha$-decay events. There are two important points on this plot: with a 99\% BRR, we have SA $\approx$ 75\%, and with a 99.9\% BRR, we have SA $\approx$ 48\%. These results do not take systematic uncertainties into account.

\section{Results}
Using 300,000 simulated events of each class for training and testing the classifier, the best result of the BDT model is an accuracy score of $\approx$ 93\% and a signal acceptance of $\approx$ 48\%, with a background rejection rate of 99.9\%. Already we have a result that shows significant improvement in the mitigation of backgrounds from $\alpha$-decays in the neck pending the evaluation of systematic uncertainties. Increasing the number of simulated events and combining this type of classifier with other types (random forest, multilayer perceptron, etc.), it is possible that for the same background rejection, the signal acceptance may be even higher. 

This approach will be used to mitigate alpha-decays from the neck of DEAP-3600 experiment. Based on the outcome we plan to consider stacking method - combining several classifiers into one large model in order to reduce the error of individual classifier. We also plan to increase Monte Carlo statistics to further improve the analysis.

\bibliographystyle{unsrt}  
\bibliography{references}  

\begin{thebibliography}{1}

\bibitem{chen2016xgboost}
Tianqi Chen and Carlos Guestrin.
\newblock Xgboost: A scalable tree boosting system.
\newblock In {\em Proceedings of the 22nd acm sigkdd international conference
  on knowledge discovery and data mining}, pages 785--794, 2016.

\bibitem{amaudruz2019design}
P-A Amaudruz, M~Baldwin, M~Batygov, B~Beltran, CE~Bina, D~Bishop, J~Bonatt,
  G~Boorman, Mark~Guy Boulay, B~Broerman, et~al.
\newblock Design and construction of the deap-3600 dark matter detector.
\newblock {\em Astroparticle Physics}, 108:1--23, 2019.

\bibitem{collaboration2019search}
DEAP Collaboration, R~Ajaj, P-A Amaudruz, GR~Araujo, M~Baldwin, M~Batygov,
  B~Beltran, CE~Bina, J~Bonatt, MG~Boulay, et~al.
\newblock Search for dark matter with a 231-day exposure of liquid argon using
  deap-3600 at snolab.
\newblock {\em Physical Review D}, 100(2):022004, 2019.

\bibitem{albertsson2018machine}
Kim Albertsson, Piero Altoe, Dustin Anderson, John Anderson, Michael Andrews,
  Juan Pedro~Araque Espinosa, Adam Aurisano, Laurent Basara, Adrian Bevan,
  Wahid Bhimji, et~al.
\newblock Machine learning in high energy physics community white paper.
\newblock {\em arXiv preprint arXiv:1807.02876}, 2018.

\bibitem{murphy2012machine}
Kevin~P Murphy.
\newblock {\em Machine learning: a probabilistic perspective}.
\newblock MIT press, 2012.

\end{thebibliography}

\end{document}